%% file: main.tex
\documentclass[sigconf,authorversion,nonacm]{acmart}
\usepackage{tikz}
\usepackage{enumerate}
\usepackage{multirow}
\usepackage{array}
\usepackage{listings}
\usepackage[ruled,vlined]{algorithm2e}
\usepackage{amsmath}
\usepackage{amsfonts}
\usepackage{pifont}
\usepackage{hyperref}
\usepackage{float}
\usepackage[utf8]{inputenc}
\usepackage{makecell}
\usepackage[inference]{semantic}
\usepackage{algorithm2e}

\newcommand{\code}[1]{\allowbreak\texttt{\small{#1}}}
\definecolor{gitadd}{HTML}{00A64F}
\definecolor{gitdel}{HTML}{c94238}
\newcommand{\diffadd}[1]{\color{gitadd}{\texttt{#1}}}

\newcommand{\xmark}{\ding{55}}

\newcommand{\bugcount}{\text{14}}

\AtBeginDocument{%
  \providecommand\BibTeX{{%
    \normalfont B\kern-0.5em{\scshape i\kern-0.25em b}\kern-0.8em\TeX}}}

\copyrightyear{2022} 
\acmYear{2022} 
\setcopyright{usgovmixed}\acmConference[ASE '22]{37th IEEE/ACM International Conference on Automated Software Engineering}{October 10--14, 2022}{Rochester, MI, USA}
\acmBooktitle{37th IEEE/ACM International Conference on Automated Software Engineering (ASE '22), October 10--14, 2022, Rochester, MI, USA}
\acmPrice{15.00}
\acmDOI{10.1145/3551349.3556937}
\acmISBN{978-1-4503-9475-8/22/10}

\begin{document}

\title{SA4U: Practical Static Analysis for Unit Type Error Detection}

\author{Max Taylor}
\email{taylor.2751@osu.edu}
\affiliation{%
  \institution{The Ohio State University}
  \country{United States}
}

\author{Johnathon Aurand}
\email{aurand.15@osu.edu}
\affiliation{%
  \institution{The Ohio State University}
  \country{United States}
}

\author{Feng Qin}
\email{qin.34@osu.edu}
\affiliation{%
  \institution{The Ohio State University}
  \country{United States}
}

\author{Xiaorui Wang}
\email{wang.3596@osu.edu}
\affiliation{%
  \institution{The Ohio State University}
  \country{United States}
}

\author{Brandon Henry}
\email{brandon.henry@tangramflex.com}
\affiliation{
\institution{Tangram Flex}
\country{United States}
}

\author{Xiangyu Zhang}
\email{xyzhang@cs.purdue.edu}
\affiliation{%
  \institution{Purdue University}
  \country{United States}
}


\begin{abstract}
\input{abstract}
\end{abstract}




\maketitle

\input{i_introduction}

\input{ii_nutshell}

\input{iii_system_overview}

\input{iv_implementation}

\input{v_evaluation}

\input{vi_casestudies}

\input{vii_related_work}

\input{viii_conclusion}

\bibliographystyle{ACM-Reference-Format}
\bibliography{acmart.bib}

\end{document}

%% file: abstract.tex
Unit type errors, where values with physical unit types (e.g., meters, hours) are used incorrectly in a computation, are common in today's unmanned aerial system (UAS) firmware. Recent studies show that unit type errors represent over 10\% of bugs in UAS firmware. Moreover, the consequences of unit type errors are severe. Over 30\% of unit type errors cause UAS crashes. This paper proposes SA4U: a practical system for detecting unit type errors in real-world UAS firmware. SA4U requires no modifications to firmware or developer annotations. It \emph{deduces} the unit types of program variables by analyzing simulation traces and protocol definitions. SA4U uses the deduced unit types to identify when unit type errors occur. SA4U is effective: it identified \bugcount{} previously undetected bugs in two popular open-source firmware (ArduPilot \& PX4.) 

%% file: i_introduction.tex
\section{Introduction}
\label{sec:introduction}
Unit type errors (UTEs) occur when developers mistakenly use incorrect physical units in a computation. For example, developers may accidentally store a value that represents a physical quantity measured in centimeters into a variable that is meant to store a value measured in meters. This occurs because the types of program variables (e.g., \lstinline{int}, \lstinline{double}) do not convey the physical units they are expected to store.

Unfortunately, UTEs are common in unmanned aerial systems (UAS) firmware. Recent work has shown that UTEs account for over 10\% of bugs in UAS firmware \cite{TaylorICUAS21}. Moreover, the consequences of UTEs are severe: Over 30\% cause UAS crashes. There are several infamous disasters caused by UTEs: a \$125 million spacecraft was lost when developers failed to convert between the metric system and imperial units, and a satellite was unavailable for one week, among others \cite{NasaOrbiter} \cite{SatelliteLost}.

A unit type has two components: a \emph{dimension} (e.g., distance and time) and a \emph{frame of reference}. Dimensions are measured in physical units such as meters and hours. Frames of reference describe how a measurement was obtained. For example, a distance is relative to a starting location, which could be the front of the vehicle or somewhere else.

Today, developers have two choices in automated methods to handle UTEs in their source code. First, they may use a \emph{unit library} such as C++'s BoostUnits \cite{BoostUnits} or Java's CaliperSharp \cite{CaliperSharp}. A unit library wraps primitive types (e.g., \lstinline{int}) in a semantic wrapper (e.g., \lstinline{Meter}).
Similarly, developers may choose to implement checks using the pattern demonstrated in \cite{allen04}. Second, developers may use a dimensional analysis tool (e.g., Phys \cite{Phys18}, PhysFrame \cite{PhysFrame21}, and \cite{Ore17}) to detect conversion errors.

There is a major drawback associated with unit libraries: Developers must annotate all program variables with their unit types. This puts a burden on developers and offers little help for existing firmware. Moreover, unit libraries either introduce runtime overhead (e.g., CalipherSharp) or only check limited expressions (e.g., BoostUnits). Today, unit libraries are unused in popular open-source UAS firmware. 

While addressing the aforementioned problems, existing tools using dimensional analysis have their own limitations. Phriky-units \cite{phriky} helps reduce the burden on developers whose projects use Robotic Operating System (ROS) \cite{ROS}. Phriky-units annotates ROS's APIs, uses type inference to propagate the unit types of program variables, and then checks that the unit types are used correctly. But the annotation burden for developers who do not use ROS is huge. Phys \cite{Phys18} reduces the annotation burden by extracting unit type information from variable names. However, large code bases often use inconsistent units internally. For example, a variable called \lstinline{altitude} could store a value measured in centimeters or meters. PhysFrame \cite{PhysFrame21} extends Phriky-units and Phys to also check the reference system of measurements. Once again, developers who do not use ROS assume an annotation burden. Moreover, prior work only considers the \emph{dimension} of a measurement (e.g., distance and time). These approaches fail to recognize when imperial and metric units are mixed. 

These facts motivate us to develop SA4U (pronounced ``safe for you''): \textbf{S}tatic \textbf{A}nalysis for \textbf{U}AS. SA4U analyzes C and C++ programs to detect UTEs. The status quo of firmware development and existing solutions motivates us to design SA4U to:
\begin{enumerate}
    \item \textbf{Work without developer annotations}: research suggests annotation burden remains a dominant factor in the lack of adoption of existing work, despite the state-of-the-art's effort to reduce annotation burden \cite{McKeever21}.
    \item \textbf{Check units}: existing tools only analyze the dimension of a measurement, and not its specific unit.
    \item \textbf{Check reference frames}: in our experience, reference frames are often harder for developers to reason about than unit types.
\end{enumerate}

To work without developer annotations, SA4U obtains type information from two sources: (1) protocol files, and (2) program traces. Protocol files define messages exchanged between the pilot's control computer and the UAS. Common protocols (MAVLink \cite{MAVLink} and LMCP \cite{LMCP}) distribute files that define the unit types of fields in messages. These files are used by existing firmware to generate data structures. We find that they are good sources for type checking too.

It is not always possible to propagate unit types from protocol files to all program variables because some expressions cannot be typed precisely (e.g., \lstinline{vector} and \lstinline{set}). To handle this challenge, SA4U uses program traces to deduce the types of program variables. SA4U instruments firmware to sample writes to program variables. Inspired by the success of simulations in related work (e.g., \cite{Timperley18} and \cite{Avis21}), SA4U executes the firmware in a simulation to obtain a trace file. SA4U's type deduction engine analyzes the trace file to identify the unit types of program variables.

The types obtained from the type deduction engine and protocol files are provided to SA4U. SA4U parses the firmware's source code and applies a set of inference rules to generate a set of constraints. SA4U
feeds these constraints to the Z3 theorem prover \cite{z3}, then reports a UTE if the constraints are unsatisfiable.

SA4U partially interprets the source code as it parses to handle scenarios where the value of a variable controls the unit type of another variable, as is common with communication protocols such as MAVLink. SA4U over-approximates possible runtime values of these so-called \emph{control variables} to precisely diagnose UTEs. SA4U keeps its run-time manageable by only approximating the runtime values of control variables.

To summarize, our contributions are:
\begin{itemize}
    \item \textbf{A type deduction engine} that mines data from firmware simulations to deduce both unit types and reference frames.
    \item \textbf{SA4U}, a practical prototype for detecting UTEs in real-world source code. We also created a prototype editor plugin to provide developers with online feedback. 
    \item \textbf{Experimental results} on two popular open-source firmware: ArduPilot \& PX4. Our results show it is possible to accurately perform type deduction on program variables in UAS. They also show SA4U is effective, identifying \bugcount{} previously undetected UTEs.
\end{itemize}

%% file: ii_nutshell.tex
\section{SA4U in a Nutshell}

This section presents three key challenges that existing approaches face when detecting UTEs. Then we show how SA4U's new unit type representation combined with its mining-assisted static analysis approach addresses these challenges.

\subsection{Three Key Challenges}

\begin{figure}
    \centering
    \begin{lstlisting}
float closest_z(Location l, Location ol, Velocity v, Velocity ov, u32 time) {
    // Velocity is measured in m/s, so delta_vel_d stores m/s.
    (*@\label{fig:Example_Unit_UTE:vel_d}@*)float delta_vel_d = ov.z - v.z;
    
    // pos is measured in cm, so delta_pos_d stores cm.
    (*@\label{fig:Example_Unit_UTE:pos_d}@*)float delta_pos_d = ol.z - l.z;
    
    // ERROR: cm - m.
    (*@\label{fig:Example_Unit_UTE:ret}@*)return fabsf(delta_pos_d - delta_vel_d * time) / 100.0f;
}\end{lstlisting}
    \caption{APM-20286: A previously reported UTE.}
    \label{fig:Example_Unit_UTE}
\end{figure}

\smallskip
\noindent
\textbf {Challenge 1: Imprecise Unit Types.}
Figure \ref{fig:Example_Unit_UTE} shows a procedure from ArduPilot that calculates the expected distance along the z axis between the UAS and an obstacle after \lstinline{time} seconds. The procedure computes how fast the vehicle is approaching the obstacle at line \ref{fig:Example_Unit_UTE:vel_d}. Then, it calculates the distance between the vehicle and the obstacle at line \ref{fig:Example_Unit_UTE:pos_d}. Finally, the procedure returns the final distance along the z axis at line \ref{fig:Example_Unit_UTE:ret}. However, line \ref{fig:Example_Unit_UTE:ret} contains a subtle UTE. Velocity is measured in meters per second, but position is measured in centimeters. Thus, line \ref{fig:Example_Unit_UTE:ret} mistakenly subtracts meters from centimeters. This error causes the firmware to fail to recognize that the UAS will pass too close to an obstacle.

Prior work fails to diagnose the UTE in Figure \ref{fig:Example_Unit_UTE}. Phys represents physical unit types as $b_i^j$, where $b_i$ is a base unit in one dimension defined by the International System of Units (SI) \cite{SIBaseUnits} and $j \in \mathbb{Z}$. So, the type of \lstinline{delta_pos_d} is represented as \lstinline{meters^1}. The type of \lstinline{delta_vel_d} is represented as \lstinline{meters^1 * seconds^-1}. The computation at line \ref{fig:Example_Unit_UTE:ret} is therefore considered legal, since subtracting meters from meters is allowed. Phys cannot detect 
UTEs that involve different measurement units in the same dimension due to this design decision.

This type of error is likely to occur in practice. MAVLink \cite{MAVLink}, measures distance in meters in over 450 message fields. However, it also measures distance in centimeters in 35 other message fields. As another example, time is measured in microseconds in over 100 message fields, but is also measured in milliseconds in 60 others. It is easy to forget to convert between units within the same dimension. Tooling is needed to catch this type of error.

\begin{figure}
    \centering
    \begin{lstlisting}
(*@\label{fig:Instrumented_Source_Code:1}@*)void handle_obstacle_distance_msg(const mavlink_obstacle_distance_t &msg) {
  ...
(*@\label{fig:Instrumented_Source_Code:patch}@*)(*@\diffadd{+\ \ if (msg.frame != MAV\_FRAME\_BODY\_FRD) \{ }@*)
(*@\diffadd{+\ \ \ \ log("Unsupported frame"); }@*)
(*@\label{fig:Instrumented_Source_Code:ret}@*)(*@\diffadd{+\ \ \ \ return; }@*)
(*@\diffadd{+\ \ \} }@*)
  (*@\label{fig:Instrumented_Source_Code:2}@*)set_obstacle_boundary(msg.angle, msg.distance / 100.0);
  ...
}

void set_obstacle_boundary(float angle, float distance) {
  ...
  (*@\label{fig:Instrumented_Source_Code:angle_assgn}@*)_angle = angle;
  (*@\label{fig:Instrumented_Source_Code:dist_assgn}@*)_distance = distance;
  ...
}\end{lstlisting}
    \caption{APM-16903: A previously undetected UTE discovered by SA4U.}
    \label{fig:Instrumented_Source_Code}
\end{figure}



\smallskip
\noindent
\textbf {Challenge 2: Missing Frames of Reference.}
Figure \ref{fig:Instrumented_Source_Code} shows a simplified version of ArduPilot's message handler for obstacle distances. A separate system onboard the UAS detects obstacles and communicates their distances to the firmware with this message. Execution of the firmware's message handler starts at line \ref{fig:Instrumented_Source_Code:1}. The handler performs some processing of the message and then invokes the \lstinline{set_obstacle_boundary} function at line \ref{fig:Instrumented_Source_Code:2} with the message's \lstinline{angle} and \lstinline{distance} fields. The body of \lstinline{set_obstacle_boundary} assigns these fields to the \lstinline{_angle} and \lstinline{_distance} fields at lines \ref{fig:Instrumented_Source_Code:angle_assgn} and \ref{fig:Instrumented_Source_Code:dist_assgn}. This is a UTE, since the program variable \lstinline{_angle} is measured in the vehicle's reference frame (i.e., \lstinline{MAV_FRAME_BODY_FRD}). Meanwhile, \lstinline{msg}'s members could be measured in \emph{any} reference frame, e.g., relative to the front of the obstacle sensor instead of the vehicle's reference frame. This error can cause the UAS to fail to recognize that its current trajectory leads to a collision, since \lstinline{_angle}'s frame of reference is managed incorrectly. After consulting with ArduPilot's developers, we created the patch shown at line \ref{fig:Instrumented_Source_Code:patch}. ArduPilot does not mean to support coordinate systems other than the vehicle's body frame for this message, so the patch simply discards messages with unexpected frames.

Both Phys and Phriky do not consider a measurement's frame of reference, so they are unable to identify the UTE in Figure \ref{fig:Instrumented_Source_Code}. PhysFrame was designed to catch errors involving frames of reference. But it targets systems that use ROS, and cannot be trivially modified for use on this system.


\begin{figure}
    \centering
    \begin{lstlisting}
vector<mav_mission_item> waypoints;

void handle_waypoint(mav_mission_item &m) {
    // Logic to validate m not shown.
    ...
    waypoints.append(m);
}\end{lstlisting}
    \caption{ArduPilot's procedure to handle waypoints.}
    \label{fig:Container_Listing}
\end{figure}

\smallskip
\noindent
\textbf {Challenge 3: Inaccurate Unit Typing via Static Inference.}
One possible way to infer unit types for program variables is based on pure static inference, as Phriky does. Specifically, one can start from the unit types that are precisely defined in the communication protocols between the firmware and the UAS or the controllers (e.g., MAVLink in ArduPilot and PX4), then propagate the unit types using dataflow analysis, and finally detect inconsistency of unit types using dimensional analysis. However, the presence of complex variable types (e.g., \lstinline{vector}, \lstinline{set}) in C++ prevents the propagation of unit types to many variables in the source code of firmware via static analysis. 

\autoref{fig:Container_Listing} shows such an example, a simplified code snippet from ArduPilot. This procedure handles waypoints a user uploads for the UAS to navigate to. The unit type of the fields in \lstinline{mav_mission_item} can vary, depending on the message's frames of reference. As a result, the container variable \lstinline{waypoints} could have many elements with different unit types. When assigning an element out of this container to a variable \lstinline{nextwaypoint}, a static inference tool must conservatively assign all possible unit types to the variable, which easily leads to inference explosion.

Due to similar reasons, prior work has shown that only a small fraction of variables can be assigned with unique unit types in ROS projects~\cite{Phys18}. To address this challenge, Phys leverages the hints from variable names as a source of unit types. However, variable names in UAS firmware often provide misleading information about the unit type of a variable. For example, Phys uses the rule that variables whose name ends with \lstinline{position} stores values in meters. But this is not always true in UAS firmware.

\subsection{How does SA4U Help?}
SA4U introduces three critical ideas. First, SA4U improves the precision of the representation of unit types. Specifically, SA4U differentiates between types \emph{in the same dimension.} So, meters and centimeters are different types. Second, SA4U enhances the unit type representation with the frames of reference so that it can detect the inconsistency between variables with different frames of reference. Third, SA4U infers likely types of program variables through profiled values at run time, in addition to the unit types defined in the message fields in the communication protocol files.

\smallskip
\noindent \textbf {SA4U's Unit Type Representation.}
A \emph{unit type} in SA4U is defined as a tuple $(Unit,~ Frame)$. In contrast to prior work, SA4U encodes precise measurement unit information in its unit representation. SA4U represents the unit of a measurement as \lstinline{s * $b_i^j$}, where $b_i$ is an SI unit \cite{SIBaseUnits} (meter, second, mole, ampere, kelvin, candela, or gram) and $s \in \mathbb{R}$. $s$ is the log10 of the measurement unit's scalar multiple, so that type checking is decideable. All physical units (e.g., volts) can be expressed as a combination of SI base units. The frame is a constant (e.g., \lstinline{MAV_FRAME_BODY_FRD}) or \lstinline{ANY} that identifies the frame of reference for the measurement.

\smallskip
\noindent\textbf{How does SA4U address challenge 1?} Consider the bug shown in Figure \ref{fig:Example_Unit_UTE}. SA4U represents the unit of \lstinline{delta_vel_d} as \lstinline{0 * meters * seconds^-1.} The type of \lstinline{time} is \lstinline{seconds}. However, the type of \lstinline{delta_pos_d} is represented as \lstinline{-2 * meters}. SA4U reports an error at line \ref{fig:Example_Unit_UTE:ret} since \lstinline{delta_vel_d * time} and \lstinline{delta_pos_d} are subtracted, but their types are different.

This unit type representation also allows SA4U to elegantly distinguish imperial and metric units. It enables SA4U to diagnose UTEs like the one mentioned in \S \ref{sec:introduction}. For example, SA4U represents yards as \lstinline[keepspaces]{log10(0.91) * meters}. In contrast, prior work proposes handling this case by introducing a separate unit type for each imperial unit. But this direction is not practical: False positives would be reported even if developers correctly converted between unit systems since the representation of different units is orthogonal.

\smallskip
\noindent\textbf{How does SA4U address challenge 2?} 
SA4U constrains the frames of variables based on the definition it extracts from protocol files. Consider the example shown in Figure \ref{fig:Instrumented_Source_Code} without the patch. SA4U learns \lstinline[keepspaces=true]{msg.frame = frame(msg.angle)} from MAVLink's protocol file. SA4U initially approximates the value of \lstinline{msg.frame} to be \lstinline{Any}. Since \lstinline{_angle} is assigned with \lstinline{msg.angle}, and \lstinline[keepspaces=true]{frame(_angle) != Any}, SA4U reports the mismatch error. 

Now, consider the behavior of SA4U with the patch. SA4U witnesses the conditional statement at line \ref{fig:Instrumented_Source_Code:patch}, so SA4U refines its approximation of the value \lstinline{msg.frame} as follows. After observing the return statement at line \ref{fig:Instrumented_Source_Code:ret}, SA4U applies the complement of the refined estimate in the remaining function body. Thus, SA4U recognizes that \lstinline[keepspaces=true,breakatwhitespace=true]{frame(mavlink_obstacle_distance_t.angle) = MAV_FRAME_BODY_FRD.} SA4U does not report a false positive in the patched version.

\smallskip
\noindent \textbf{How does SA4U address challenge 3?} SA4U mines likely unit types from program traces obtained from simulated UAS execution. Specifically, SA4U inserts instrumentation to sample the runtime values of program variables. SA4U also samples the physical states (e.g., position, acceleration, velocity) of objects in the simulation. SA4U knows the unit types of simulated objects a priori (with minimal annotations from us). SA4U compares the values of program variables with the simulation's physical states to deduce the variable's likely unit type. In total, SA4U samples 15 values per simulated object. These values correspond to entries in the MAVLink \lstinline{HIL_STATE_QUATERNION} message (i.e., the minimum information required to perform a simulation). This instrumentation is only required once per \emph{simulator}, and allows all firmware that support the simulator to benefit from SA4U.

For example, ArduPilot contains a variable called \lstinline[breakatwhitespace=true]{target_altitude.} This variable stores the altitude that the user commanded the UAS to navigate to. SA4U compares \lstinline{target_altitude} with the simulation's physical state, including the UAS' current altitude stored in the field \lstinline{altitude}. SA4U deduces \lstinline{target_altitude} and \lstinline{altitude} share the same unit type, since the UAS' altitude is always eventually approximate to the value of
\lstinline{target_altitude}.

%% file: iii_system_overview.tex
\section{SA4U Design}

\begin{figure}
    \centering
    \includegraphics{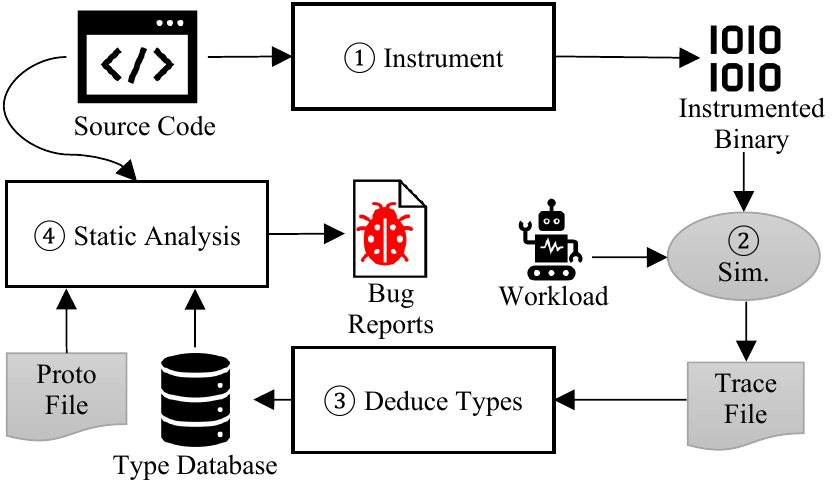}
    \caption{SA4U's workflow. The numbered boxes show each stage of SA4U's execution. Components shown in gray are not part of SA4U. Icons show each stage's inputs and outputs.}
    \label{fig:SA4U_Overview}
\end{figure}

Figure \ref{fig:SA4U_Overview} shows a high-level overview of SA4U. Developers run SA4U's instrumentation program on firmware source code to insert instrumentation that tracks the runtime values of program variables. Then, developers run the instrumented binary in a simulator. In this step, the instrumentation creates a trace file that contains the runtime values of program variables. The instrumentation also writes quantities of interest (i.e. the values of physical quantities with known unit types) to the trace file. Next, the type deduction program mines the trace file to produce the type database. The type database tracks the types of some program variables. Finally, the static analysis stage employs unit type information from the type database and protocol file, and conducts type inference on the firmware source code to check types are used correctly.



\subsection{Instrumentation}
\label{subsec:instrumentation}

    

SA4U's instrumentation tool inserts code to record the values of the firmware's program variables. One potential problem with instrumentation is the disturbance of the timing during the firmware execution. As a result, 
the firmware detects starvation and triggers fail-safes that prevent the workload from executing normally. To prevent this scenario, we exploit the following two observations:
\begin{enumerate}
    \item \textbf{We do not need all program variables.} 
    The types of local variables can often be inferred from the types of non-local variables used within the same function based on their interactions. Therefore, we only instrument the accesses (loads and stores) of variables with non-local lifetimes.
    \item \textbf{We do not need all the values of a program variable.} The instrumentation limits the number of values recorded to a fixed rate (in our case, 1 per second). Since the same values of a variable could be stored up to tens of thousands of times each second, this greatly reduces the amount of data that must be recorded and analyzed.
\end{enumerate}

The instrumentation tool builds a partial function $ID: \text{Names} \rightarrow \mathbb{N}$ that maps variable names to integer IDs. When the instrumentation tool encounters an unseen name $n$, it defines $ID(n)$ to be the next available ID. The instrumentation tool outputs the definition of $ID$ for subsequent phases of SA4U.


\subsection{Type Deduction}
\label{subsec:Type_Deduction}

\begin{table*}[h]
    \centering
    \begin{tabular}{c|c}
        \hline 
        \textbf{Rule} & \textbf{Intuition} \\
        \hline
        $q_1: U,~ \square var_1 \approx q_1 \models var_1: U$ & \makecell{Always approximately equal values share a type.} \\
        $q_1: U,~ \square var_1 \approx C_0 \times q_1  \models var_1: C_0 \times U$ & \makecell{Values that share a linear relation share a type.} \\
        $q_1: U,~ \square var_1 \approx x \implies \lozenge q_1 \approx x \models var_1: U \text{ (and the reverse)}$ & \text{Prophecies tell types.} \\
        \hline
    \end{tabular}
    \caption{Type deduction rules used by SA4U.}
    \label{tab:Type_Deduction_Rules}
\end{table*}

The goal of the type deduction stage is to assign unit types to the program variables that appear in the trace file produced by the simulation stage. To accomplish this, we derive the unit types from several \emph{quantities of interest}. Quantities of interest are measurements with types known a priori (e.g., vehicle altitude in meters). We trace the values of quantities of interest in the simulator. Then, we check if one of several relationships hold between program variables and each quantity of interest to deduce types. 

To be precise, the input to the type deduction stage is a set of observations $T$ obtained from the simulation stage, and the $ID$ function from the instrumentation stage. An observation is a tuple $(\mathbb{Z}^+,~ \mathbb{N},~ \mathbb{R})$, where the first component is a timestamp, the second component is a unique variable ID (i.e. from the $ID$ function), and the third component is an observed value.

Before performing type deduction, we apply several filters to program variables to reduce 
false positives. First, we remove all program variables whose values are a constant $< 10$ or whose static type is an \lstinline{enum}. Small constants are unlikely to be related to physical quantities, and enums definitely are not.  Second, we remove variables that are only written once. This is because the written-once variables are not related to quantities of interest, which represent ever-changing aspects of the environment.

Table \ref{tab:Type_Deduction_Rules} uses linear temporal logic (LTL) to describe the rules SA4U uses to assign types to some (but not all) of the program variables that appear in $T$. In the table, $var_1$ and $var_2$ refer to arbitrary variable IDs that appear in $T$, and $q_1$ refers to an arbitrary quantity of interest. In LTL, the $\square$ operator means ``always'' and the $\lozenge$ operator means ``eventually.'' Notice that our rules accept \emph{approximate} equality. We say that $var \approx q$ if $|var - q| / |q| < \epsilon.$ We permit approximate equality because we do not track every value written to both $var$ and $q$. Section \ref{subsubsec:sensitivity_analysis} discusses the sensitivity of our rules with respect to $\epsilon$.

To summarize the rules in Table \ref{tab:Type_Deduction_Rules} in plain-speak:
\begin{enumerate}
    \item Rule 1 says that if $var_1$ is always approximately $q_1$, then $var_1$ and $q_1$ share a unit type. This allows SA4U to deduce the type of \lstinline{_angle} in Figure \ref{fig:Instrumented_Source_Code}, since \lstinline{_angle}'s value is close to the angle between the UAS and its closest obstacle. 
    \item Rule 2 says that if $var_1$ has a linear dependence on $q_1$, then the linear dependence can be applied across their types.
    \item Rule 3 says that if $var_1$ predicts a value that a quantity $q_1$ eventually takes on, then $var_1$ and $q_1$ share a type. This allows SA4U to learn types like \lstinline{target_altitude: m} since the UAS always reaches its target altitude.
\end{enumerate}

The type deduction stage applies its rules in the order shown in Table \ref{tab:Type_Deduction_Rules}. As types are mined, a type database is assembled. The type database relates variable IDs to a unit type. Formally, the type database is a mapping $DB: \mathbb{N} \rightarrow Type$. In the case of rules 1 and 3, the type is the same as the associated quantity of interest's. In the case of rule 2, the type is named by the mined linear relationship.

\subsection{Static Analysis for Type Inference and UTE Detection}
\label{subsec:Static_Analysis}

SA4U's static analysis stage uses \emph{protocol files} that describe messages exchanged between the user's control computer and the UAS for type information. Static analysis also uses the type database generated in the type deduction stage. Combined, this allows SA4U to infer the types of a large number of program variables. 

Before discussing the static analysis component, we need to introduce our definitions of the subtype relation $ \sqsubseteq $ and the operator $op \in \left\{+, -, \times, /\right\}$ between two unit types.
\begin {enumerate}
\item Let unit types $U_1 = (D_1,~ F_1)$ and $U_2 = (D_2,~ F_2)$. We say that $U_1 \sqsubseteq U_2 \iff (D_1 = D_2) \land (F_1 = F_2 \lor F_2 = Any)$. 

\item Consider $U_1 = (D_1, F)$ and $U_2 = (D_2, F),$ where $D_1 = c_1 * b_{1}^{i_{1,1}} * \ldots * b_{n}^{i_{1,n}}$ and $D_2 = c_2 * b_{1}^{i_{2,1}} * \ldots * b_{n}^{i_{2,n}}.$ We define $U_1~ \times~ U_2 = ((c_1 + c_2) * (b_1^{i_{1,1} + i_{2, 1}}) * \ldots * (b_n^{i_{1,n} + i_{2, n}}), F)$. We use a similar definition for division.

\item Consider $U_1 + U_2$. Without loss of generality, assume that $U_1 \sqsubseteq U_2$. We define $U_1 + U_2 = U_2$. We use a similar definition for subtraction.
\end{enumerate}

\begin{figure}
    \centering
    \begin{lstlisting}
<msg id="103" name="VISION_SPEED_ESTIMATE">
  <description>Speed estimate.</description>
  <field type="frame"
         name="frame">
    Frame
  </field>
  <field name="usec" 
         units="us">
    Timestamp
  </field>
  <field name="x" 
         units="m/s">
    Global X speed
  </field>
</msg>\end{lstlisting}
    \caption{Part of MAVLink's protocol file.}
    \label{fig:Proto_File_Example}
\end{figure}

\begin{table*}[h]
    \centering
    \begin{tabular}{c|c}
        \hline 
        \textbf{Name} & \textbf{Rule} \\
        \hline
        & \\
        Binary Operator Judgement &
        $\inference{\Gamma \vdash \code{e}_1: \code{U}_1, \code{e}_2: \code{U}_2}{\Gamma \vdash \code{e}_1~ op~ \code{e}_2: \code{U}_1~ op~ \code{U}_2}$ \\
        & \\
        Assignment Judgement  &
        $\inference{\Gamma \vdash \code{e}: \code{U}_1, \code{v}: \code{U}_2 & \vdash \code{U}_1 \sqsubseteq \code{U}_2}{\Gamma \vdash \code{v} = \code{e} : \code{U}_2, \Gamma\{\code{v} \rightarrow \code{U}_2\}}$ \\
        & \\
        Conditional Refinement  &
        $\inference{\Gamma,\Sigma \vdash \code{var}_1 = \code{val} \implies \code{var}_2: \code{U} & \vdash \overline{\code{s}} : \Gamma\{\code{var}_2 \rightarrow U\}, \Sigma}
        {\Gamma,\Sigma \vdash \code{if (var\textsubscript{1} == val)~ } \overline{\code{s}} : \Gamma\{\code{var}_2 \rightarrow U\}}$ \\
        & \\
        Variable Type Inference &
        $\inference{\Gamma,\gamma \vdash \code{T}: \code{U}}{\Gamma,\gamma \vdash \code{T v} : \code{U}, \Gamma\{v \rightarrow \code{U}\}}$ \\
        & \\
        Argument Type Inference &
        $\inference{\Gamma \vdash \code{e}_1 : \code{U}_1, \ldots, \code{e}_n : \code{U}_n}{\Gamma \vdash f(\code{e}_1, \ldots, \code{e}_n) \implies  \code{ArgType(f, 1)} = U_1, \ldots \code{ArgType}(f, n) = U_n}$\\
        & \\
        Return Type Inference &
        $\inference{\Gamma \vdash \code{ReturnType(f) = U}}{\Gamma \vdash f(\ldots): U}$ \\
        & \\
       \hline
    \end{tabular}
    \caption{The type inference rules used by SA4U in the static analysis stage.}
    \label{tab:Type_Inference_Rules}
\end{table*}

Note that we add rather than multiply the scalar coefficients of units in the definition of the $\times$ and $/$ operators. This is because we represent scalar coefficients as the $\log_{10}$ of the actual scalar coefficient. For example, we represent the unit \lstinline{centimeter} as $-2 * meter$ instead of $\frac{1}{100} * meter.$ We avoid multiplying scalar coefficients because the type checker would need to solve a system of non-linear equations. This is Hilbert's $10^{\text{th}}$ problem, which is famously undecidable. Since we restrict the scalar coefficient of the unit component to be a rational number, checking if the scalar coefficients are compatible is decidable in linear time. This is because the problem is equivalent to deciding a system of linear Diophantine equations. We select the $\log_{10}$ representation of scalar coefficients because most practical units have a scalar coefficient that is a power of ten.

\subsubsection{Protocol Files}
\label{subsec:Protocol_Files}
Figure \ref{fig:Proto_File_Example} shows an example message from MAVLink's protocol file. Firmwares such as ArduPilot and PX4 often contain generators that process the protocol file and create structs 
to represent each message. For example, \lstinline{mavlink_vision_sp-eed_estimate_t} in Figure \ref{fig:Proto_File_Example}. Message fields (e.g., \lstinline{usec}) are struct members. Protocol files precisely define hundreds of messages and thousands of fields, and thus serve as a great source of unit types.

We use a simple scheme to represent compound data types as variables. If a struct \lstinline{s} has a member \lstinline{a}, we treat \lstinline{s.a} as its own variable. We make a simplified assumption that all elements of an array share a type. So, we treat an array access \lstinline{array[i]} as an access to the variable \lstinline{array}. This helps us perform limited analysis on pointers and support arrays with variable lengths. 

SA4U needs protocol files because its type deduction stage cannot learn the types of protocol messages. This is unfortunate, since MAVLink handlers are a dominant source of UTEs. Figure \ref{fig:Proto_File_Example} illustrates the problem: the type of \code{x} depends on the value of the field \code{frame}. We introduce the constraint set $\Sigma$ to account for this kind of relationship. $\Sigma$ is a set of relationships in the template $\code{var}_1 = \code{v}_1 \implies \code{var}_2 : U.$ In the case of the example in Figure \ref{fig:Proto_File_Example}, $\Sigma$ contains the relationship \lstinline{vision_speed_estimate.frame} = $ GLOBAL \implies$ \lstinline{vision_speed_estimate.x} $: (1\times m \times s^{-1}, GLOBAL).$

Protocol files provide two pieces of information used in the static analysis stage. The first piece of information is $\gamma$, which relates program variable types (e.g., \lstinline{vision_speed_estimate_t}) to unit types. The second piece of information is the set of control relationships $\Sigma$.

\subsubsection{Type Inference and UTE Detection}
The inputs to the type inference system are the stack-frame model $\Gamma$, the relationship between program types and unit types $\gamma$, the control relationship between variables $\Sigma$, and the type database \code{DB}. $\Gamma$ tracks the types of variables in the current stack-frame. Table \ref{tab:Type_Inference_Rules} describes the type inference rules SA4U applies in the static analysis stage. 

The binary operator judgement rule allows SA4U to perform dimensional analysis on expressions involving multiple types. Informally, it reads that if the current stack frame $\Gamma$ assigns the type $U_1$ to $e_1$ and $U_2$ to $e_2$, then the type of the expression $e_1~ op~ e_2$ is $U_1~ op~ U_2.$ For example, \lstinline[keepspaces]{x: (m, GLOBAL) / t: (s, GLOBAL)} is assigned the type \lstinline{(m * s^-1, GLOBAL)} according to this rule. Note that SA4U reports a type error when a binary operator cannot be typed according to this rule. For example, SA4U will report a UTE error if it finds \lstinline {x: (m: GLOBAL) + t: (s, GLOBAL)}.

The assignment judgement rule prevents illegal stores to program variables. Informally, it says that if an expression $e$ has the type $U_1$ in the current stack frame ($\Gamma$,) and the expression $v$ has the type $U_2$ in $\Gamma$, then \lstinline{e = v} can be assigned the type $U_2$ only if $U_1 \sqsubseteq U_2$. The unit type of a variable is initially unconstrained, and the frame is \lstinline{Any}. This allows gradual type refinement through assignment. Similar to the case of the binary operator judgement rule, SA4U reports an error if an assignment expression is untypeable. 

The conditional refinement rule helps SA4U use type information from protocol files. Since protocol files often use \emph{control fields} in messages to indicate the frames of other message members, we must partially interpret the source code to approximate the values that may be present in control fields. The rule in Table \ref{tab:Type_Inference_Rules} shows how SA4U uses information in conditional branches to refine the conditional type. Informally, it says that if there is a control relationship between $var_1$ and $var_2$, and the conditional statement's body $\overline{s}$ can be typed using the control relationship, then a conditional statement with suitably established value of $var_1$ infers that $\overline{s}$ can be typed. Note that we do not consider the myriad of ways that a conditional statement could establish the value of $var_1$. Instead, we observe that developers use control relationships in simple ways. They compare the value of control fields directly to constants in either if statements or switch statements. 

The variable type inference rule uses the unit types extracted from the protocol files to type program variables. Informally, it says that if a static type (i.e., a type in the actual program) $T$ has the unit type (i.e., the type used in dimensional analysis) $U$, then witnessing the declaration of a variable with type $T$ updates the stack model $\Gamma$ with $v$ assigned $U$. This allows SA4U to handle the static types from entries in the protocol file.

The argument type inference rule checks type consistency across multiple call sites. We build a free function called \lstinline{ArgTypes} that assigns a type to argument $i$ of a function $f$. If we can find a model of \lstinline{ArgTypes} then argument types are consistently provided throughout the source code. To reduce false positives, we ignore calls to functions in the standard library, and maintain a small list of function types to ignore.

Finally, the return type inference rule checks type consistency in return values between multiple call sites. Similar to argument type inference, we use a free function to model the return type of procedures. If return types are used in inconsistent ways then the constraint system will be unsatisfiable.

SA4U applies the inference rules shown in Table \ref{tab:Type_Inference_Rules} to each function in the firmware's source code to generate a type-model of the program. Specifically, when SA4U encounters an expression that matches the expression on the top of a proof bar in the table, SA4U generates the constraint shown on the bottom of the proof bar. After SA4U has parsed the entire source code, SA4U invokes the Z3 theorem prover \cite{z3} to check that the constraints are satisfiable. If they are satisfiable, then the type rules are satisified by the subject source code. Otherwise, SA4U generates a bug report that summarizes the issue.




%% file: iv_implementation.tex
\section{Implementation}

This section discusses the implementation. SA4U contains three tools: one for instrumentation, one for type deduction, and one for static analysis. We also discuss workloads and simulation. All of SA4U is open-source, and is available at \url{https://github.com/obicons/sa4u}.

\subsection{Instrumentation}

SA4U's instrumentation tool is a Clang tool \cite{clangTool} that rewrites assignment expressions. Clang tools provide a library to use the front end of the Clang C++ compiler, simplifying the task of source code transformer development. Assignment expressions are rewritten to invoke an instrumentation macro. The instrumentation macro logs the values of program variables to a CSV file.


\subsection{Simulation}

Step 2 of the SA4U's workflow requires the developer to execute the firmware in a physics simulator. 
We chose to use Gazebo \cite{gazebo} because it is robust and both ArduPilot and PX4 support it. Any simulator works for this job, but small code changes are required to track quantities of interest. 

Users must provide a workload during simulation. A ideal workload contains enough complex behavior so that the simulation will cover most of the firmware code. 
In practice, obtaining suitable workloads is straight-forward. For example, both ArduPilot and PX4 contain end-to-end functional tests that we used as a workload. 

\subsection{Type Deduction}

We implemented SA4U's type deduction tool in Python. Initially, we tried to use Daikon \cite{daikon}, a popular invariant detector. However, we discovered that Daikon's execution took far too long (on the order of days) even with our effort on restricting outputs. This is mainly because Daikon mines general invariants that SA4U does not need.  So we created a simple invariant miner for type deduction.

\subsection{Static Analysis}

Our static analysis engine is a Python tool that uses \lstinline{libclang} \cite{libclang}. \lstinline{libclang} provides simple bindings to the Clang compiler's front end. This allows our static analysis tool to analyze any code that Clang can compile. Similar to the instrumentation engine, the static analysis tool depends on widely available compilation databases. We use the Z3 theorem prover \cite{z3} to check the inference rules shown in Table \ref{tab:Type_Inference_Rules}. Z3 is an efficient formula solver, and is a popular choice for implementations similar to ours. The static analysis engine accounts for the majority of our implementation burden.

\subsection{Editor Plugin}

\begin{figure}
    \centering
    \includegraphics[scale=.25]{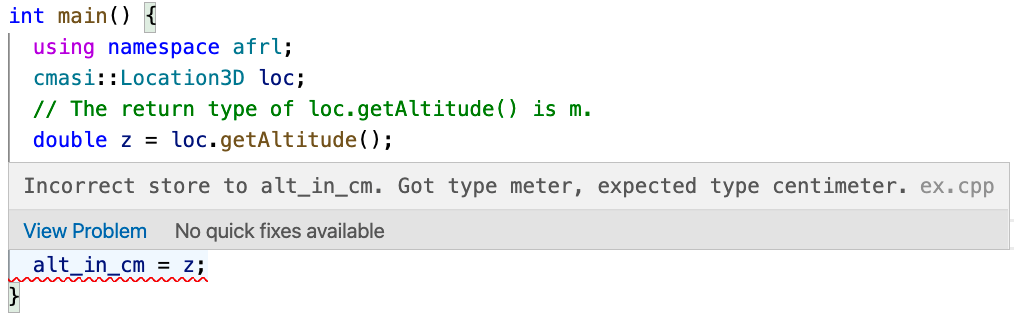}
    \caption{Screenshot of the SA4U Editor Plugin. SA4U reports an error based on the contents of the protocol file and the type database.}
    \label{fig:VSCode_Ext}
\end{figure}

We created a prototype editor plugin to make SA4U easier to use. Figure \ref{fig:VSCode_Ext} shows the SA4U editor plugin in action. We wrote a language server protocol (LSP) server that communicates with the developer's text editor. Almost all text editors (even Emacs and Vim) can use LSP servers. We tested our LSP server with a VSCode plugin. Our LSP server is a thin wrapper around the SA4U analysis binary. The server simply runs SA4U and scans its output for issues. Overall, our implementation was painless. We only had to write 150 lines of code to create the server and extension.

%% file: v_evaluation.tex
\section{Evaluation}

Our evaluation addresses three research questions: (1) How effective is SA4U at detecting UTEs in UAS firmware? (2) How does SA4U compare to the state-of-the-art? (3) How efficient is SA4U? 

\begin{table*}[ht]
    \centering
    \begin{tabular}{cc|c|c|c|c|c}
         \hline
          & & \multicolumn{3}{c|}{\textbf {UTE Features}} & \multicolumn{2}{c}{\textbf {Found by}} \\
          \cline{3-7}
         \multicolumn{2}{c|}{\textbf{Bug ID}} &
         \textbf{Same dimension?} & \textbf{Frame?} & \textbf{MAV?} & \textbf{SA4U} & \textbf{Phys}  \\
         \hline
         & APM-16903 & No  & Yes & Yes & \checkmark & \xmark  \\ 
         & APM-19868 & No  & Yes & Yes & \checkmark & \xmark \\ 
         & APM-21291 & No  & Yes & Yes & \checkmark & \xmark \\ 
         & APM-21309 & No  & Yes & Yes & \checkmark & \xmark \\
         & PX4-17337 & No  & Yes & Yes & \checkmark & \xmark \\
         & PX4-17354 & No  & Yes & Yes & \checkmark & \xmark \\
         & PX4-19973 & No  & Yes & Yes & \checkmark & \xmark  \\
         & PX4-19983 & No  & Yes & Yes & \checkmark & \xmark \\
         & PX4-19995 & No  & Yes & Yes & \checkmark & \xmark \\
         & PX4-20000 & No  & Yes & Yes & \checkmark & \xmark  \\
         & PX4-20018 & No  & Yes & Yes & \checkmark & \xmark  \\
         & PX4-20019 & No  & No & Yes & \checkmark & \xmark   \\
         & PX4-GPSDrivers-110 & Yes & No  & No & \checkmark & \xmark  \\
         & PX4-20023 & Yes  & Yes & No & \checkmark &  \xmark \\
         \hline
    \end{tabular}
    \caption{Previously unknown bugs detected by SA4U and state-of-the-art tool Phys.}
    \label{tab:Bugs_Detected}
\end{table*}

\subsection{Experiment Methodology}

\subsubsection{Platform and Target Firmware.}
We conducted our experiments on an Ubuntu 18.04 server equipped with an Intel Core i5-2500 CPU and 8GB of memory. We evaluated SA4U on ArduPilot \cite{ArduPilot} and PX4 \cite{PX4}. We chose these systems for several reasons. First, ArduPilot and PX4 are the most popular open source UAS firmware. Combined, they have more stars on Github than any other UAS project. Second, UAS firmware is a complex category of control software, whose safety is paramount. There are many onboard sensors, complex control equations, and remote message systems. UTEs in any component could lead to disastrous scenarios. Finally, we want to show that SA4U works independently of the firmare's hardware abstraction mechanisms. 

\subsubsection{Workload for Type Deduction.}
We used the same workload for ArduPilot and PX4 in the type deduction phase. The workload guides the UAS to takeoff, navigates to 8 waypoints, and then lands. We obtained this workload from ArduPilot's existing test files.

\subsubsection{SA4U's Effectiveness.} We applied SA4U to our target systems to evaluate how effectively it detects both known and unknown bugs. For known bugs, we used 6 bugs from a recent bug study \cite{TaylorICUAS21} to see whether SA4U can detect them. For unknown bugs, we evaluated SA4U on the latest versions of ArduPilot and PX4.

We compared SA4U with the state-of-the-art tool Phys for detecting unknown bugs. To make Phys work on ArduPilot and PX4, we used SA4U's code to annotate MAVLink variables. Additionally, we evaluated the false positive rates of SA4U using 5 random files from ArduPilot and PX4.

\subsubsection{SA4U's Efficiency and Sensitivity.} We conducted experiments to evaluate SA4U's execution time when running it with the latest versions of ArduPiolot and PX4. The execution time include SA4U's analysis and Z3's type inference. Also we compare SA4U's execution time with Phys's execution time over the same firmware. Furthermore, we evaluated the sensitivity of SA4U's type deduction engine on the approximation threshold $\epsilon$.

\subsection{Experimental Results}
\begin{table}[h]
    \centering
    \begin{tabular}{c|c|c}
        \hline
         \textbf{Bug ID} & \textbf{Frame?} & \textbf{Found by SA4U?} \\
         \hline
         APM-3542 & Yes  & \checkmark \\
         APM-4105 & Yes  & \checkmark \\
         APM-4550 & Yes  & \xmark \\
         PX4-12517 & Yes  & \checkmark \\
         PX4-12532 & Yes  & \checkmark \\
         PX4-13180 & Yes  & \checkmark \\
        \hline
    \end{tabular}
    \caption{Known bugs used to evaluate SA4U.}
    \label{tab:Known_Bugs}
\end{table}

\subsubsection{Detecting Known Bugs.} Table \ref{tab:Known_Bugs} shows the detection result of SA4U on 6 known bugs from a recent bug study \cite{TaylorICUAS21}. We selected these 6 bugs because other bugs were found in old versions of the firmware that we could not easily build in our environment. SA4U must be able to build the firmware in order to analyze it due to its dependency on libclang. Note that it is impossible to report a traditional recall statistic because the number of UTEs in the evaluation subjects is unknown. In the table, the bug ID column identifies the bug in the repository's bug management system. The frame column shows if a bug involves a frame of reference. SA4U was able to diagnose 5 of the 6 bugs. APM-4550 was not diagnosed because SA4U incorrectly infers that a conversion of reference frame takes place when a variable is multiplied, however the multiplication was not a conversion.

\subsubsection{Detecting Unknown Bugs.} Table \ref{tab:Bugs_Detected} shows the previously unknown bugs detected by SA4U. The unknown UTEs reported by SA4U have various features, errors within a dimension or between different dimensions, errors involving different frames of reference or within the same frame, and errors involving MAVLink message handlers. This is mainly because SA4U's interprocedural analysis and its type representation precisely capturing the unit information within a dimension and the frame information of the unit.

We noticed that most of the UTEs listed in Table \ref{tab:Bugs_Detected} span multiple procedures. We believe this is because developers often manually test new code or code revision,
and even simple manual tests are likely to expose UTEs contained in a single function. Moreover, developers can easily remember the types of local variables while they write a procedure.

However, interprocedural flows are tricky. Developers are likely to misunderstand function contracts because they are not formally documented. This leads to UTEs. Another possible reason is that developers could refactor a function and violate its contract, thus breaking previously working code. Finally, developers must keep up with the latest changes to protocol specifications (e.g., MAVLink often adds new message fields). If firmware falls behind, the values sent by pilot control computers will be misinterpreted by the firmware. For example, the bug shown in Figure \ref{fig:PX4_17354_Case_Study} was caused by developers failing to notice a change in MAVLink's specification.

\subsubsection{Comparing with the state-of-the-art.} We compared SA4U with the state-of-the-art tool in this space: Phys \cite{Phys18}. SA4U was able to find bugs in the subject systems that Phys could not detect for two reasons. First, because Phys does not consider the reference frame in its type representation. Observe that the reference frame was involved in 13 of the 14 bugs. Second, Phys is unable to detect 2 of the 14 bugs because it lacks the precise unit type information that SA4U uses. For example, in PX4-GPSDrivers-110, a driver fails to convert from milliseconds to microseconds.

Direct comparison with PhysFrame is not possible because PhysFrame is designed specifically for ROS programs. However, PhysFrame struggles with the bugs shown in Table \ref{tab:Bugs_Detected} at a conceptual level. Notice that 12 of the 14 bugs involve MAVLink. SA4U introduces the \emph{conditional refinement} rule in \autoref{tab:Type_Inference_Rules} to assign types to expression whose type depends on the value of a control field. Since PhysFrame lacks a corresponding rule, it conceptually struggles to detect these bugs.


\subsubsection{False Positives and False Negatives.} To evaluate false positives, we looked at the outputs of SA4U for 5 random files. SA4U found 32 errors in the 5 files. While this seems high, there are duplicates in the reports. For example, if a bug occurs in a function defined in a header file and the function is called in multiple source files, then there will be multiple bug reports. Among those 32 errors, we can confirm that there are 20 true errors (including duplicates). This yields a false positive rate of approximately $38\%$. 

There are multiple sources of these false positives. First, the type deduction phase can derive an incorrect type for a program variable. For example, the program variable \code{emergency\_mode\_alt} could be assigned the type centimeter, when the actual type is meter. When this occurs, SA4U's static analysis phase reports a false positive. Second, the static analysis phase itself is not sound. For example, developers use conversion functions to transform the types of values. SA4U is not aware of all conversion functions, so it misdiagnoses values whose types that are transformed this way.

It is worth noting that SA4U is neither sound (i.e., SA4U has false positives) nor complete (i.e., SA4U has false negatives). Once again, the type deduction phase could incorrectly label the type of a program variable. If this occurs, the static analysis phase could fail to diagnose a UTE. Moreover, the static analysis phase can only detect errors if a sufficiently large number of types have been inferred. Finally, sometimes programmers temporarily store values with incorrect types in variables. SA4U reports false positives in this case.

\subsubsection{SA4U's Efficiency.} Table \ref{tab:runtime} shows SA4U's execution time on its evaluation subjects. SA4U is efficient to detect UTEs in large code bases. For example, it took SA4U about 37 minutes to scan 848,562 lines of code in 7,320 files in ArduPilot.

SA4U has a speedup of up to 2.16x over the state-of-the-art Phys. Note that SA4U's analysis is almost identical to Phys'. This speedup is mainly attributed to implementation differences in SA4U. Specifically, SA4U uses Z3, a high-performance theorem prover, to perform type inference. Z3 performs type inference in parallel, thus giving SA4U a nice speedup.

\begin{table}[H]
    \centering
    \begin{tabular}{c|c|c|c}
         \hline
         \textbf{UTE Tool} & \textbf{Firmware} & \textbf{LoC} & \textbf{Runtime (seconds)} \\
         \hline
         \multirow{2}{*}{SA4U} & ArduPilot & 848,562 & 2215 \\
         & PX4 & 197,795 & 689 \\
         \hline
         \multirow{2}{*}{Phys} & ArduPilot & 848,562 & 4792 \\
         & PX4 & 197,795 & 3951\\
         \hline
    \end{tabular}
    \caption{Runtime of UTE detection tools.}
    \label{tab:runtime}
\end{table}

\subsubsection{Invariant Mining Sensitivity}
\label{subsubsec:sensitivity_analysis}
\begin{table}
    \centering
    \begin{tabular}{c|c|c|c}
    \hline
    \textbf{$\epsilon$} & \textbf{Approximate} & \textbf{Linear} & \textbf{Eventually} \\
    \hline
    100\% & 16823 & 343 & 58 \\
    97.5\% & N/A & 1331 & 8613 \\
    95\% & N/A & 1569 & 19841 \\
    90\% & 3625 & 2248 & 32754 \\
    80\% & 3610 & 3707 & 41287 \\
    70\% & 3594 & 5916 & 53003 \\
    60\% & 3501 & 9097 & 63867 \\
    50\% & 3493 & 14539 & 87498 \\
    40\% & 3470 & 20857 & 107657 \\
    30\% & 3463 & 32131 & 130617 \\
    20\% & 3444 & 51887 & 159957 \\
    10\% & 3444 & 90694 & 252289 \\
    5\% & 3382 & N/A & 318665 \\
    2.5\% & 3373 & N/A & 322034 \\
    1\% & 3363 & N/A & 322056 \\
    \hline
    \end{tabular}
    \caption{The effect of $\epsilon$ on type deduction sensitivity. \textbf{``Approximate''} refers to rule 1 in Table \ref{tab:Type_Deduction_Rules}, \textbf{``Linear''} refers to rule 2, and \textbf{``Eventually''} refers to rule 3.}
    \label{tab:Sensitivity_Table}
\end{table}

Table \ref{tab:Sensitivity_Table} shows the effect of the choice of $\epsilon$ on the number of types mined by the type deduction engine. $\epsilon$'s meaning is overloaded for each invariant template. Recall that $\epsilon$ allows us to tolerate the measurement errors we introduced by only periodically sampling the values of variables. In the case of \textbf{Approximate}, $\epsilon$ controls the relative error between the measurements. So, as the permissible relative error becomes smaller, fewer invariants are mined. In the case of \textbf{Linear}, $\epsilon$ controls the absolute value of the Pearson correlation coefficient where we accept the relationship. As the Pearson correlation approaches 0, the relationship between two variables is less linear. Finally, in the case of \textbf{Eventually}, $\epsilon$ controls the confidence threshold where the mined invariant is accepted.

We used the values in Table \ref{tab:Sensitivity_Table} to select $\epsilon$ when we ran SA4U's type deduction engine. We selected $\epsilon = 5\%$ for \textbf{Approximate} invariants because the extra mined relationships were useful versus the $2.5\%$ and $1\%$ levels. For \textbf{Linear} and \textbf{Eventually} invariants we used 0.975.

%% file: vi_casestudies.tex
\section{Case Studies}

\begin{figure}
    \centering
    \begin{lstlisting}
void on_vsn_spd_est(vision_speed_estimate_t e) {
  ...
  (*@\label{fig:APM_19868:call}@*)e.usec = corrector.correct_time(
    e.usec, 
    lcl_time()
  );
  ...
}
    
// Accounts for transport delay for the
// timestamp rmt received at time local.
(*@\label{fig:APM_19868:handler}@*)int Corr::correct_time(int rmt, int lcl) {
  ...
  (*@\label{fig:APM_19868:diff}@*)int diff = lcl - link_offset;
  (*@\label{fig:APM_19868:instr}@*)est_rmt_time = diff;
  (*@\label{fig:APM_19868:link_offset}@*)link_offset = est_rmt_time - rmt;
  ...
}\end{lstlisting}
    \caption{APM-19868: A UTE affecting ArduPilot.}
    \label{fig:APM_19868}
\end{figure}

We discuss the following two UTE cases reported by SA4U in more details to illustrate its detection capability.

\subsubsection{APM-19868}
\begin{sloppypar}
Figure \ref{fig:APM_19868} shows a UTE that occurred in ArduPilot. MAVLink defines the \lstinline{vision_speed_estimate_t} message to communicate the estimated speed of the UAS based on the input from an external visual navigation system. The \lstinline{vision_speed_estimate_t} message contains the field \lstinline{usec} that communicates the time when the estimate was generated. This field is measured in microseconds, 
storing either the time since the system booted, or the time since the UNIX epoch (i.e. January 1, 1970). The receiver is expected to use the magnitude of the field to determine the timestamp's format. 
\end{sloppypar}

\begin{sloppypar}
Line \ref{fig:APM_19868:handler} of Figure \ref{fig:APM_19868} shows ArduPilot's handler for the \lstinline{vision_speed_estimate_t} message. SA4U's instrumentation program adds instrumentation to line \ref{fig:APM_19868:instr} since it contains a store to a non-local variable. Next, we simulate the UAS flight to obtain a trace file. Then, the type deduction program analyzes the values stored in \lstinline{est_rmt_time}. The quantity of interest that \lstinline{est_rmt} best matches is microseconds  since system boot. This type information is recorded to the type database.
\end{sloppypar}

Then, SA4U runs the static analysis stage and parses the source code that contains the functions shown in Figure \ref{fig:APM_19868}. When parsing line \ref{fig:APM_19868:call}, SA4U knows that the type of \code{e.usec} can be in either UNIX epoch format (i.e. \lstinline{TIME_UNIX}) or time since system boot (i.e. \lstinline{TIME_BOOT}) because of the contents of the protocol file. So, SA4U generates the constraint \lstinline[keepspaces]|ArgType(Corr::correct_time, 1) = (1e-6  * s, {TIME_BOOT, TIME_UNIX})|. Later, When encountering line \ref{fig:APM_19868:diff}, SA4U applies its binary operator judgement rule and generates the constraints \lstinline[keepspaces]{type(local) = type(link_offset)} and \lstinline[keepspaces]{type(diff) = type(lcl)}. Next, SA4U generates the constraint \lstinline[keepspaces]{type(est_rmt_time) = type(diff)} when it parses line \ref{fig:APM_19868:instr}. Notice that this simplifies to \lstinline[keepspaces]{type(est_rmt_time) = type(lcl)}. Finally, SA4U once again applies the binary operator judgement rule to generate the constraint that \lstinline[keepspaces]{type(est_rmt_time) = ArgType(Corr::correct_time, 1)} when it parses line \ref{fig:APM_19868:link_offset}. But this constraint is unsatisfiable due to a conflict between the constraints from the type database and first constraint SA4U introduced:  \lstinline[keepspaces]|ArgType(Corr::correct_time, 1) = (1e-6  * s, {TIME_BOOT, TIME_UNIX})|. Since the constraint cannot be satisfied, SA4U creates a UTE bug report.

We reported this bug to the developers of ArduPilot. They confirmed that this is indeed a UTE, however no patch has been accepted at the time of writing. Developers are still unsure about the best way to change \lstinline{Corr::correct_time} to remove the bug.

\begin{figure}
    \centering
    \begin{lstlisting}
ImageTargetHandler handler;
void on_landing_target(landing_target &t) {
    if (t.has_pos && t.frame == LOCAL) {
        handle_pos_target(t);
(*@\diffadd{+\ \ \ \ \} else if (t.has\_pos) \{ }@*)
(*@\diffadd{+\ \ \ \ \ \ \ \ \ log("Unsupported frame.");}@*)    
    } else {
        handler.set_target(t);
    }
}

void ImageTargetHandler::set_target(landing_target &t) {
    ...
    _x = t.img_angle_x;
    _y = t.img_angle_y;
}\end{lstlisting}
    \caption{PX4-17354: A complex UTE SA4U found in PX4.}
    \label{fig:PX4_17354_Case_Study}
\end{figure}

\subsubsection{PX4-17354} 
Figure \ref{fig:PX4_17354_Case_Study} shows a UTE that occurred in PX4. The \lstinline{LANDING_TARGET} MAVLink message communicates a landing location to the UAS. This message supports two different uses. First, pilots can command the UAS to land in a location in a reference picture. Second, pilots can command the UAS to land at a  position specified as GPS coordinates. In the second case, the landing target can be specified in any coordinate system.

Line 4 of Figure \ref{fig:PX4_17354_Case_Study} checks if the coordinate system of the landing target is local since PX4 only supports the local coordinate system for landing messages. However, if the frame is not local, the else branch on line 8 is taken. This causes read accesses to two image fields (\lstinline{img_angle_x} and \lstinline{img_angle_y}) of \lstinline{t} at lines 15 and 16, respectively. However, the message may contain a landing position specified in GLOBAL coordinate system. Thus, the two image fields at lines 15 and 16 are undefined and these accesses are incorrect.

\begin{sloppypar}
SA4U identifies this bug, albeit somewhat crudely. First, SA4U correctly learns the types of \lstinline{ImageTargetHandler::_x} and \lstinline{ImageTargetHandler::_y}. Then, SA4U performs static analysis. In the \lstinline{else} branch on line 8 of Figure \ref{fig:PX4_17354_Case_Study}, SA4U represents the coordinate system of each of \lstinline{t}'s members as any possible coordinate system, minus local. Then, SA4U sees the function call on line 9. This call must be incorrect since the possible frames of \lstinline{t} are not compatible with the stores at lines 15 and 16. 
\end{sloppypar}

We patched the bug by introducing the code shown on lines 6 and 7. When we reported the bug and submitted our patch to PX4's developers, they confirmed the bug and accepted our change.

%% file: vii_related_work.tex
\section{Related Work}
\textbf{Dimensional Analysis.} Dimensional analysis is a widely used technique to validate equations. Early work provided programming language support and packages, e.g. \cite{karr}. Osprey \cite{osprey} used dimensional analysis to validate C programs, but required manual annotations. Later work (e.g. \cite{Phys18}, \cite{PhysFrame21}, \cite{Ore17}) tried to reduce annotation burden several ways. Phys \cite{Phys18} uses variable names to infer unit types. PhysFrame \cite{PhysFrame21} extends Phys to also consider the frame of the unit. PhysFrame is a well-designed tool, but it is built specifically for programs that use ROS. It cannot be trivially applied to the systems studied here. Phriky-units \cite{phriky}  and \cite{Ore17} reduces the annotation burden by pre-annotating shared libraries, and then using type inference to deduce the types of program variables. 

\textbf{Invariant Mining.} Mining program invariants is a mature idea. Daikon \cite{daikon} is the most prolific example. Later work (e.g. \cite{temporalInvariants}) extends invariant mining to capture temporal invariants, an idea we use in our eventually-equal mining rule.  Researchers often use mined invariants in program analysis. For example, CoFi \cite{cofi} mines invariants in distributed system and injects faults at points where invariants do not hold. Similar to our work, MonkeyType \cite{MonkeyType} runs Python unit tests to discover likely program variable types.

%% file: viii_conclusion.tex
\section{Conclusion}
We presented SA4U, a tool for finding UTEs in real UAS firmware. SA4U obtains type information from traces of program variables, and protocol files. SA4U partially interprets the source code of UAS to constrain the unit types of program variables. This allows SA4U to analyze message handlers of common protocols. Then, SA4U applies dimensional analysis to infer the types of variables not known from other sources. In the future, we wish to extend SA4U in two ways. First, we plan to ease developer burden by developing repair tools to generate patches for UTEs. Second, we hope to integrate other source of type information (e.g., Phys) with SA4U.

\begin{acks}
The authors would like to thank the anonymous reviewers for their valuable feedback and thoughtful suggestions. This work is partially sponsored by the grants: AFRL FA864921P0206, NSF 1901242, and ONR N000142012733.
\end{acks}

%% file: main.bbl

\begin{thebibliography}{29}


\ifx \showCODEN    \undefined \def \showCODEN     #1{\unskip}     \fi
\ifx \showDOI      \undefined \def \showDOI       #1{#1}\fi
\ifx \showISBNx    \undefined \def \showISBNx     #1{\unskip}     \fi
\ifx \showISBNxiii \undefined \def \showISBNxiii  #1{\unskip}     \fi
\ifx \showISSN     \undefined \def \showISSN      #1{\unskip}     \fi
\ifx \showLCCN     \undefined \def \showLCCN      #1{\unskip}     \fi
\ifx \shownote     \undefined \def \shownote      #1{#1}          \fi
\ifx \showarticletitle \undefined \def \showarticletitle #1{#1}   \fi
\ifx \showURL      \undefined \def \showURL       {\relax}        \fi
\providecommand\bibfield[2]{#2}
\providecommand\bibinfo[2]{#2}
\providecommand\natexlab[1]{#1}
\providecommand\showeprint[2][]{arXiv:#2}

\bibitem[\protect\citeauthoryear{??}{Ard}{2021}]%
        {ArduPilot}
 \bibinfo{year}{2021}\natexlab{}.
\newblock \bibinfo{title}{ArduPilot}.
\newblock \bibinfo{howpublished}{\url{https://ardupilot.org}}.
\newblock


\bibitem[\protect\citeauthoryear{??}{Cal}{2021}]%
        {CaliperSharp}
 \bibinfo{year}{2021}\natexlab{}.
\newblock \bibinfo{title}{CaliperSharp}.
\newblock
  \bibinfo{howpublished}{\url{https://github.com/point85/CaliperSharp}}.
\newblock


\bibitem[\protect\citeauthoryear{??}{MAV}{2021}]%
        {MAVLink}
 \bibinfo{year}{2021}\natexlab{}.
\newblock \bibinfo{title}{MAVLink}.
\newblock \bibinfo{howpublished}{\url{https://mavlink.io}}.
\newblock


\bibitem[\protect\citeauthoryear{??}{PX4}{2021}]%
        {PX4}
 \bibinfo{year}{2021}\natexlab{}.
\newblock \bibinfo{title}{PX4: Open Source Autopilot for Drone Developers}.
\newblock \bibinfo{howpublished}{\url{https://px4.io}}.
\newblock


\bibitem[\protect\citeauthoryear{??}{ROS}{2021}]%
        {ROS}
 \bibinfo{year}{2021}\natexlab{}.
\newblock \bibinfo{title}{ROS - Robot Operating System}.
\newblock \bibinfo{howpublished}{\url{https://www.ros.org}}.
\newblock


\bibitem[\protect\citeauthoryear{??}{cla}{2022}]%
        {clangTool}
 \bibinfo{year}{2022}\natexlab{}.
\newblock \bibinfo{title}{Clang Tools}.
\newblock
  \bibinfo{howpublished}{\url{https://clang.llvm.org/docs/ClangTools.html}}.
\newblock


\bibitem[\protect\citeauthoryear{??}{lib}{2022}]%
        {libclang}
 \bibinfo{year}{2022}\natexlab{}.
\newblock \bibinfo{title}{libclang}.
\newblock
  \bibinfo{howpublished}{\url{https://clang.llvm.org/doxygen/group__CINDEX.html}}.
\newblock


\bibitem[\protect\citeauthoryear{Allen, Chase, Luchangco, Maessen, and
  Steele~Jr}{Allen et~al\mbox{.}}{2004}]%
        {allen04}
\bibfield{author}{\bibinfo{person}{Eric Allen}, \bibinfo{person}{David Chase},
  \bibinfo{person}{Victor Luchangco}, \bibinfo{person}{Jan-Willem Maessen},
  {and} \bibinfo{person}{Guy~L Steele~Jr}.} \bibinfo{year}{2004}\natexlab{}.
\newblock \showarticletitle{Object-oriented units of measurement}. In
  \bibinfo{booktitle}{\emph{Proceedings of the 19th annual ACM SIGPLAN
  conference on Object-oriented programming, systems, languages, and
  applications}}. \bibinfo{pages}{384--403}.
\newblock


\bibitem[\protect\citeauthoryear{Beschastnikh, Brun, Ernst, Krishnamurthy, and
  Anderson}{Beschastnikh et~al\mbox{.}}{2011}]%
        {temporalInvariants}
\bibfield{author}{\bibinfo{person}{Ivan Beschastnikh}, \bibinfo{person}{Yuriy
  Brun}, \bibinfo{person}{Michael~D. Ernst}, \bibinfo{person}{Arvind
  Krishnamurthy}, {and} \bibinfo{person}{Thomas~E. Anderson}.}
  \bibinfo{year}{2011}\natexlab{}.
\newblock \showarticletitle{Mining Temporal Invariants from Partially Ordered
  Logs}. In \bibinfo{booktitle}{\emph{Managing Large-Scale Systems via the
  Analysis of System Logs and the Application of Machine Learning Techniques}}
  (Cascais, Portugal) \emph{(\bibinfo{series}{SLAML '11})}.
  \bibinfo{publisher}{Association for Computing Machinery},
  \bibinfo{address}{New York, NY, USA}, Article \bibinfo{articleno}{3},
  \bibinfo{numpages}{10}~pages.
\newblock
\showISBNx{9781450309783}
\urldef\tempurl%
\url{https://doi.org/10.1145/2038633.2038636}
\showDOI{\tempurl}


\bibitem[\protect\citeauthoryear{{Bureau International des Poids et
  Mesures}}{{Bureau International des Poids et Mesures}}{2019}]%
        {SIBaseUnits}
\bibfield{author}{\bibinfo{person}{{Bureau International des Poids et
  Mesures}}.} \bibinfo{year}{2019}\natexlab{}.
\newblock \bibinfo{title}{{The International System of Units (SI)}}.
\newblock
  \bibinfo{howpublished}{\url{https://www.bipm.org/documents/20126/41483022/SI-Brochure-9-EN.pdf/2d2b50bf-f2b4-9661-f402-5f9d66e4b507}}.
\newblock


\bibitem[\protect\citeauthoryear{Chen, Dou, Wang, and Qin}{Chen
  et~al\mbox{.}}{2020}]%
        {cofi}
\bibfield{author}{\bibinfo{person}{Haicheng Chen}, \bibinfo{person}{Wensheng
  Dou}, \bibinfo{person}{Dong Wang}, {and} \bibinfo{person}{Feng Qin}.}
  \bibinfo{year}{2020}\natexlab{}.
\newblock \showarticletitle{CoFI: consistency-guided fault injection for cloud
  systems}. In \bibinfo{booktitle}{\emph{Proceedings of the 35th IEEE/ACM
  International Conference on Automated Software Engineering}}.
  \bibinfo{pages}{536--547}.
\newblock


\bibitem[\protect\citeauthoryear{de~Moura and Bj{\o}rner}{de~Moura and
  Bj{\o}rner}{2008}]%
        {z3}
\bibfield{author}{\bibinfo{person}{Leonardo de Moura} {and}
  \bibinfo{person}{Nikolaj Bj{\o}rner}.} \bibinfo{year}{2008}\natexlab{}.
\newblock \showarticletitle{Z3: An Efficient SMT Solver}. In
  \bibinfo{booktitle}{\emph{Tools and Algorithms for the Construction and
  Analysis of Systems}}, \bibfield{editor}{\bibinfo{person}{C.~R. Ramakrishnan}
  {and} \bibinfo{person}{Jakob Rehof}} (Eds.). \bibinfo{publisher}{Springer
  Berlin Heidelberg}, \bibinfo{address}{Berlin, Heidelberg},
  \bibinfo{pages}{337--340}.
\newblock
\showISBNx{978-3-540-78800-3}


\bibitem[\protect\citeauthoryear{Duquette}{Duquette}{[n.\,d.]}]%
        {LMCP}
\bibfield{author}{\bibinfo{person}{Matthew Duquette}.}
  \bibinfo{year}{[n.\,d.]}\natexlab{}.
\newblock \bibinfo{booktitle}{\emph{The Common Mission Automation Services
  Interface}}.
\newblock
\urldef\tempurl%
\url{https://doi.org/10.2514/6.2011-1542}
\showDOI{\tempurl}
\showeprint{https://arc.aiaa.org/doi/pdf/10.2514/6.2011-1542}


\bibitem[\protect\citeauthoryear{Ernst, Czeisler, Griswold, and Notkin}{Ernst
  et~al\mbox{.}}{2000}]%
        {daikon}
\bibfield{author}{\bibinfo{person}{Michael~D. Ernst}, \bibinfo{person}{Adam
  Czeisler}, \bibinfo{person}{William~G. Griswold}, {and}
  \bibinfo{person}{David Notkin}.} \bibinfo{year}{2000}\natexlab{}.
\newblock \showarticletitle{Quickly detecting relevant program invariants}. In
  \bibinfo{booktitle}{\emph{ICSE 2000, Proceedings of the 22nd International
  Conference on Software Engineering}}. \bibinfo{address}{Limerick, Ireland},
  \bibinfo{pages}{449--458}.
\newblock


\bibitem[\protect\citeauthoryear{Jiang and Su}{Jiang and Su}{2006}]%
        {osprey}
\bibfield{author}{\bibinfo{person}{Lingxiao Jiang} {and}
  \bibinfo{person}{Zhendong Su}.} \bibinfo{year}{2006}\natexlab{}.
\newblock \showarticletitle{Osprey: a practical type system for validating
  dimensional unit correctness of C programs}. In
  \bibinfo{booktitle}{\emph{Proceedings of the 28th international conference on
  Software engineering}}. \bibinfo{pages}{262--271}.
\newblock


\bibitem[\protect\citeauthoryear{Karr and Loveman}{Karr and Loveman}{1978}]%
        {karr}
\bibfield{author}{\bibinfo{person}{Michael Karr} {and}
  \bibinfo{person}{David~B. Loveman}.} \bibinfo{year}{1978}\natexlab{}.
\newblock \showarticletitle{Incorporation of Units into Programming Languages}.
\newblock \bibinfo{journal}{\emph{Commun. ACM}} \bibinfo{volume}{21},
  \bibinfo{number}{5} (\bibinfo{date}{may} \bibinfo{year}{1978}),
  \bibinfo{pages}{385–391}.
\newblock
\showISSN{0001-0782}
\urldef\tempurl%
\url{https://doi.org/10.1145/359488.359501}
\showDOI{\tempurl}


\bibitem[\protect\citeauthoryear{Kate, Chinn, Choi, Zhang, and Elbaum}{Kate
  et~al\mbox{.}}{2021}]%
        {PhysFrame21}
\bibfield{author}{\bibinfo{person}{Sayali Kate}, \bibinfo{person}{Michael
  Chinn}, \bibinfo{person}{Hongjun Choi}, \bibinfo{person}{Xiangyu Zhang},
  {and} \bibinfo{person}{Sebastian Elbaum}.} \bibinfo{year}{2021}\natexlab{}.
\newblock \showarticletitle{PHYSFRAME: Type Checking Physical Frames of
  Reference for Robotic Systems}. In \bibinfo{booktitle}{\emph{Proceedings of
  the 29th ACM Joint Meeting on European Software Engineering Conference and
  Symposium on the Foundations of Software Engineering}} (Athens, Greece)
  \emph{(\bibinfo{series}{ESEC/FSE 2021})}. \bibinfo{publisher}{Association for
  Computing Machinery}, \bibinfo{address}{New York, NY, USA},
  \bibinfo{pages}{45–56}.
\newblock
\showISBNx{9781450385626}
\urldef\tempurl%
\url{https://doi.org/10.1145/3468264.3468608}
\showDOI{\tempurl}


\bibitem[\protect\citeauthoryear{Kate, Ore, Zhang, Elbaum, and Xu}{Kate
  et~al\mbox{.}}{2018}]%
        {Phys18}
\bibfield{author}{\bibinfo{person}{Sayali Kate}, \bibinfo{person}{John-Paul
  Ore}, \bibinfo{person}{Xiangyu Zhang}, \bibinfo{person}{Sebastian Elbaum},
  {and} \bibinfo{person}{Zhaogui Xu}.} \bibinfo{year}{2018}\natexlab{}.
\newblock \showarticletitle{Phys: Probabilistic Physical Unit Assignment and
  Inconsistency Detection}. In \bibinfo{booktitle}{\emph{Proceedings of the
  2018 26th ACM Joint Meeting on European Software Engineering Conference and
  Symposium on the Foundations of Software Engineering}} (Lake Buena Vista, FL,
  USA) \emph{(\bibinfo{series}{ESEC/FSE 2018})}.
  \bibinfo{publisher}{Association for Computing Machinery},
  \bibinfo{address}{New York, NY, USA}, \bibinfo{pages}{563–573}.
\newblock
\showISBNx{9781450355735}
\urldef\tempurl%
\url{https://doi.org/10.1145/3236024.3236035}
\showDOI{\tempurl}


\bibitem[\protect\citeauthoryear{Koenig and Howard}{Koenig and
  Howard}{[n.\,d.]}]%
        {gazebo}
\bibfield{author}{\bibinfo{person}{Nathan Koenig} {and} \bibinfo{person}{Andrew
  Howard}.} \bibinfo{year}{[n.\,d.]}\natexlab{}.
\newblock \showarticletitle{Design and use paradigms for gazebo, an open-source
  multi-robot simulator}. In \bibinfo{booktitle}{\emph{2004 IEEE/RSJ
  International Conference on Intelligent Robots and Systems (IROS)(IEEE Cat.
  No. 04CH37566)}}, Vol.~\bibinfo{volume}{3}. IEEE,
  \bibinfo{pages}{2149--2154}.
\newblock


\bibitem[\protect\citeauthoryear{McKeever, Bennich-Bj{\"o}rkman, and
  Salah}{McKeever et~al\mbox{.}}{2021}]%
        {McKeever21}
\bibfield{author}{\bibinfo{person}{Steve McKeever}, \bibinfo{person}{Oscar
  Bennich-Bj{\"o}rkman}, {and} \bibinfo{person}{Omar-Alfred Salah}.}
  \bibinfo{year}{2021}\natexlab{}.
\newblock \showarticletitle{Unit of measurement libraries, their popularity and
  suitability}.
\newblock \bibinfo{journal}{\emph{Software: Practice and Experience}}
  \bibinfo{volume}{51}, \bibinfo{number}{4} (\bibinfo{year}{2021}),
  \bibinfo{pages}{711--734}.
\newblock


\bibitem[\protect\citeauthoryear{{Meyer, Carl}}{{Meyer, Carl}}{2017}]%
        {MonkeyType}
\bibfield{author}{\bibinfo{person}{{Meyer, Carl}}.}
  \bibinfo{year}{2017}\natexlab{}.
\newblock \bibinfo{title}{{Let your code type-hint itself: introducing open
  source MonkeyType}}.
\newblock
  \bibinfo{howpublished}{\url{https://instagram-engineering.com/let-your-code-type-hint-itself-introducing-open-source-monkeytype-a855c7284881}}.
\newblock


\bibitem[\protect\citeauthoryear{Ore, Detweiler, and Elbaum}{Ore
  et~al\mbox{.}}{2017a}]%
        {Ore17}
\bibfield{author}{\bibinfo{person}{John-Paul Ore}, \bibinfo{person}{Carrick
  Detweiler}, {and} \bibinfo{person}{Sebastian Elbaum}.}
  \bibinfo{year}{2017}\natexlab{a}.
\newblock \showarticletitle{Lightweight Detection of Physical Unit
  Inconsistencies without Program Annotations}. In
  \bibinfo{booktitle}{\emph{Proceedings of the 26th ACM SIGSOFT International
  Symposium on Software Testing and Analysis}} (Santa Barbara, CA, USA)
  \emph{(\bibinfo{series}{ISSTA 2017})}. \bibinfo{publisher}{Association for
  Computing Machinery}, \bibinfo{address}{New York, NY, USA},
  \bibinfo{pages}{341–351}.
\newblock
\showISBNx{9781450350761}
\urldef\tempurl%
\url{https://doi.org/10.1145/3092703.3092722}
\showDOI{\tempurl}


\bibitem[\protect\citeauthoryear{Ore, Detweiler, and Elbaum}{Ore
  et~al\mbox{.}}{2017b}]%
        {phriky}
\bibfield{author}{\bibinfo{person}{John-Paul Ore}, \bibinfo{person}{Carrick
  Detweiler}, {and} \bibinfo{person}{Sebastian Elbaum}.}
  \bibinfo{year}{2017}\natexlab{b}.
\newblock \showarticletitle{Phriky-Units: A Lightweight, Annotation-Free
  Physical Unit Inconsistency Detection Tool}. In
  \bibinfo{booktitle}{\emph{Proceedings of the 26th ACM SIGSOFT International
  Symposium on Software Testing and Analysis}} (Santa Barbara, CA, USA)
  \emph{(\bibinfo{series}{ISSTA 2017})}. \bibinfo{publisher}{Association for
  Computing Machinery}, \bibinfo{address}{New York, NY, USA},
  \bibinfo{pages}{352–355}.
\newblock
\showISBNx{9781450350761}
\urldef\tempurl%
\url{https://doi.org/10.1145/3092703.3098219}
\showDOI{\tempurl}


\bibitem[\protect\citeauthoryear{Sawyer}{Sawyer}{1999}]%
        {NasaOrbiter}
\bibfield{author}{\bibinfo{person}{Kathy Sawyer}.}
  \bibinfo{year}{1999}\natexlab{}.
\newblock \bibinfo{title}{Mystery of Orbiter Crash Solved}.
\newblock
  \bibinfo{howpublished}{\url{https://www.washingtonpost.com/wp-srv/national/longterm/space/stories/orbiter100199.htm}}.
\newblock


\bibitem[\protect\citeauthoryear{Schabel and Watanabe}{Schabel and
  Watanabe}{2008}]%
        {BoostUnits}
\bibfield{author}{\bibinfo{person}{Matthias Schabel} {and}
  \bibinfo{person}{Steven Watanabe}.} \bibinfo{year}{2008}\natexlab{}.
\newblock \bibinfo{title}{Boost.Units 1.0.0}.
\newblock \bibinfo{howpublished}{\url{http://boost.cowic.de/rc/pdf/units.pdf}}.
\newblock


\bibitem[\protect\citeauthoryear{Taylor, Boubin, Chen, Stewart, and Qin}{Taylor
  et~al\mbox{.}}{2021a}]%
        {TaylorICUAS21}
\bibfield{author}{\bibinfo{person}{Max Taylor}, \bibinfo{person}{Jayson
  Boubin}, \bibinfo{person}{Haicheng Chen}, \bibinfo{person}{Christopher
  Stewart}, {and} \bibinfo{person}{Feng Qin}.}
  \bibinfo{year}{2021}\natexlab{a}.
\newblock \showarticletitle{A Study on Software Bugs in Unmanned Aircraft
  Systems}. In \bibinfo{booktitle}{\emph{2021 International Conference on
  Unmanned Aircraft Systems (ICUAS)}}. \bibinfo{pages}{1439--1448}.
\newblock
\urldef\tempurl%
\url{https://doi.org/10.1109/ICUAS51884.2021.9476844}
\showDOI{\tempurl}


\bibitem[\protect\citeauthoryear{Taylor, Chen, Qin, and Stewart}{Taylor
  et~al\mbox{.}}{2021b}]%
        {Avis21}
\bibfield{author}{\bibinfo{person}{Max Taylor}, \bibinfo{person}{Haicheng
  Chen}, \bibinfo{person}{Feng Qin}, {and} \bibinfo{person}{Christopher
  Stewart}.} \bibinfo{year}{2021}\natexlab{b}.
\newblock \showarticletitle{Avis: In-Situ Model Checking for Unmanned Aerial
  Vehicles}. In \bibinfo{booktitle}{\emph{2021 51st Annual IEEE/IFIP
  International Conference on Dependable Systems and Networks (DSN)}}.
  \bibinfo{pages}{471--483}.
\newblock
\urldef\tempurl%
\url{https://doi.org/10.1109/DSN48987.2021.00057}
\showDOI{\tempurl}


\bibitem[\protect\citeauthoryear{Timperley, Afzal, Katz, Hernandez, and
  Le~Goues}{Timperley et~al\mbox{.}}{2018}]%
        {Timperley18}
\bibfield{author}{\bibinfo{person}{Christopher~Steven Timperley},
  \bibinfo{person}{Afsoon Afzal}, \bibinfo{person}{Deborah~S. Katz},
  \bibinfo{person}{Jam~Marcos Hernandez}, {and} \bibinfo{person}{Claire
  Le~Goues}.} \bibinfo{year}{2018}\natexlab{}.
\newblock \showarticletitle{Crashing Simulated Planes is Cheap: Can Simulation
  Detect Robotics Bugs Early?}. In \bibinfo{booktitle}{\emph{2018 IEEE 11th
  International Conference on Software Testing, Verification and Validation
  (ICST)}}. \bibinfo{pages}{331--342}.
\newblock
\urldef\tempurl%
\url{https://doi.org/10.1109/ICST.2018.00040}
\showDOI{\tempurl}


\bibitem[\protect\citeauthoryear{Trella, Herring, Freeman, Kilpatrick, Reth,
  Greenfield, Credland, Laine, Machi, and Smith}{Trella et~al\mbox{.}}{1998}]%
        {SatelliteLost}
\bibfield{author}{\bibinfo{person}{Massimo Trella}, \bibinfo{person}{Ellen
  Herring}, \bibinfo{person}{Richard Freeman}, \bibinfo{person}{William
  Kilpatrick}, \bibinfo{person}{Alan Reth}, \bibinfo{person}{Michael
  Greenfield}, \bibinfo{person}{John Credland}, \bibinfo{person}{Robert Laine},
  \bibinfo{person}{Dino Machi}, {and} \bibinfo{person}{Alan Smith}.}
  \bibinfo{year}{1998}\natexlab{}.
\newblock \bibinfo{title}{SOHO MISSION INTERRUPTION JOINT ESA/NASA
  INVESTIGATION}.
\newblock
  \bibinfo{howpublished}{\url{https://umbra.nascom.nasa.gov/soho/SOHO_final_report.html}}.
\newblock


\end{thebibliography}
